\documentstyle[12pt,epsf]{article}
\newcommand{\br}{\begin{array}}
\newcommand{\er}{\end{array}}
\newcommand{\beq}{\begin{equation}}
\newcommand{\eeq}{\end{equation}}

\newcommand{\G}{\Gamma}
\newcommand{\p}{\partial}

\def\ba{\begin{eqnarray}}
\def\ea{\end{eqnarray}}
\newcommand{\be}{\begin{equation}}
\newcommand{\ee}{\end{equation}}
\newcommand{\bea}{\begin{eqnarray}}
\newcommand{\eea}{\end{eqnarray}}

\makeatletter
\@addtoreset{equation}{section}
\makeatother

\def\appendix#1{
  \setcounter{section}{1}
  \setcounter{equation}{0}
  \renewcommand{\thesection}{\Alph{section}}
  \section*{Appendix \thesection\protect\indent \parbox[t]{11.715cm} {} }
  \addcontentsline{toc}{section}{Appendix \thesection\ \ \ }}

\begin{document}
\begin{flushright}
YITP-SB-00-50\\
\end{flushright}

\begin{center}
{\LARGE Propagators for p-forms in $AdS_{2p+1}$ and 
correlation functions in the $AdS_7$/(2,0) CFT  correspondence}
\vskip1truecm

{\large\bf Iosif Bena\\ 
{\it University of California, Santa Barbara CA 93106}\\
Horatiu Nastase and Diana Vaman\\
{\it C.N.Yang Institute for Theoretical Physics,\\ S.U.N.Y. Stony Brook, 
NY 11794-3840, 
USA }}\footnote{ 
E-mail addresses: iosif@physics.ucsb.edu\protect\\
hnastase@insti.physics.sunysb.edu and\protect\\
dvaman@insti.physics.sunysb.edu
} 
\vskip2truecm
\end{center}
\abstract{In $AdS_{2p+1}$ we construct propagators for p-forms whose 
lagrangians contain terms of the form $A \wedge d A$. In particular we 
explore the 
case of forms satisfying ``self duality in odd dimensions'', and the case of 
forms with a topological mass term. We point out that the ``complete'' set of 
maximally symmetric bitensors previously used in all the other propagator 
papers is incomplete - there exists another bitensor which can and does 
appear in the formulas for the propagators in this particular case. 
Nevertheless, its presence does not affect the other propagators computed 
so far.

On the $AdS$ side of the correspondence we compute the 2 and 3 point 
functions involving the self-dual tensor of the maximal $7d$ gauged 
supergravity (sugra),
$S_{\mu\nu\rho}$. 
 Since the 7 dimensional antisymmetric 
self-dual tensor obeys first order field equations ($S + * d S=0$), 
to get a nonvanishing 2 point function we add a certain boundary  term (to 
satisfy  the variational principle on a manifold with boundary) to the 
$7d$ action. The 3 point functions  we 
compute are of the type $SSB$ and $SBB$, describing vertex interactions
with the gauge fields $B_{\mu}$.
}

\newpage


\section{Introduction}

$${}$$

During the last few years, the $AdS-CFT$ correspondence has generated a lot of
interest (see \cite{agmoo} for a review). However, most of the work focused 
on the 
$AdS_5$/$SYM_4$ correspondence because of the interest in describing 
strongly coupled 4 dimensional field theories and because (non)perturbative 
results in these field theories were known. 

An interesting case of the 
correspondence is the one between string theory on $AdS_7 \times R^4$ and the 
mysterious 6 dimensional (2,0) $CFT$. This theory is a fixed point of the 
renormalization group flow from the theory on N parallel M5 branes.  
Not too much is known about this $CFT$, and so there is no independent check 
of the
$AdS/CFT$ correspondence besides verifying conformal invariance and comparing
the masses of the primaries with those obtained in the $DLCQ$ description of 
this $CFT$ \cite{agmoo}.

There are however many interesting new features in this correspondence, one of 
which is the appearance in the 7 dimensional theory of a three-index 
antisymmetric tensor 
with a  first order equation of motion (of ``self-duality in odd dimensions'' 
type - $*d A = m A$).
Finding its propagator and the corresponding $CFT$ two-point function presents 
some new challenges. The reason for studying it is twofold. 

First, by analyzing correlators of the 3-form field we learn about $CFT$ 
correlators of the self-dual 3-form in $6d$. The (2,0) $CFT$ is hard to 
describe mostly because
we don't understand how to make a self-dual 3-form nonabelian. 
The study of these 3-forms via the $AdS-CFT$ correspondence may be a 
step towards that. 
Second, two of us argued in a previous paper \cite{nv2} that one needs 
to take a nonlinear ansatz for the embedding of AdS fields into the 
higher-dimensional theory (in this case M theory) for the $AdS/CFT$ 
correspondence.
If one were able to compute correlators
independently on the $CFT$ side and compare them with the $AdS$ results 
obtained here, 
this would give further evidence for this statement.

We start by analyzing the propagator of $p$-forms in $2p+1$ dimensions, and 
for generality we look both at the case of self-duality in odd dimensions
and at the propagator for the Maxwell action with a topological mass term,
both of which are relevant in 7 dimensions.  We extend the basis of maximally
symmetric bitensors (first introduced by Allen and Jacobson \cite{allen})
with a bitensor constructed from contractions of the $\epsilon$ symbol 
with derivatives
of the $AdS$ chordal distance. We express the propagator ansatz in 
this basis, and use the equations of motion to compute it. We also present an
extra possible term in the bitensor ansatz for the propagator of $p$-forms 
in $2p$ dimensions. 

We then use the $AdS_7$ propagator for the ``self-dual'' 3-form to extract 
information about the 6 dimensional (2,0) $CFT$. The bulk to boundary 
propagator 
is obtained straightforwardly as a limit of the bulk to bulk propagator. 
Nevertheless, since the 3-form action is first order in derivatives, the 
quadratic action
vanishes on-shell. In order to use the $AdS-CFT$ correspondence, one has to 
supplement the action with a boundary
term which will generate the correct $CFT$ 2-point function. 
This procedure was first introduced for spin 1/2 fields \cite{mv,hs}, 
used in a 3-point function calculation for interacting spinors-scalars by 
\cite{gpf}, and was later 
justified by Arutyunov and Frolov \cite{af1} and Henneaux \cite{hen} by
enforcing the variational principle on a manifold with boundary.
We also compute the 3-point functions of two 3-forms and a gauge field, 
and of two gauge 
fields and a 3-form. For their calculation we follow the conformal methods
of Freedman et al \cite{fmmr}. 
The results are given in (\ref{3pjoo}) and (\ref{3pjjo}).

The paper is organized as follows. In section 2.1 we discuss the propagator
for the case of ``self-duality in odd dimensions''; in section 2.2 we 
discuss the propagator for $p$-forms with a topological mass 
term, while in the last section dedicated to propagators, 2.3, we
investigate the effect of the extra bitensors on the other propagators 
computed so far. In section 3.1 we study the 2-point function of the 3-form
field and in section 3.2 the 3-point functions of two gauge fields, and 
a 3-form and two 3-forms and a gauge field. We finish with conclusions in 
section 4. We give some useful identities involving the chordal distance 
in Appendix A.1. In Appendix A.2 we derive the limits we need when computing 
the 2-point function, while in Appendix A.3 we included some integrals used
for the 3-point functions.

\section{Propagators}
$${}$$
In the recent years a lot of papers \cite{df1,df2,fred,io,io2,asad,muck} 
have been written computing propagators of various fields in $AdS_{d+1}$. 
In the cases of tensor propagators, the standard procedure for computing 
a propagator is to express it using a basis for maximally symmetric 
bitensors (first introduced by Allen and Jacobson \cite{allen}), and to 
use the equation of motion. Typically one obtains a system of equations 
which can be solved in a straightforward way.

Nevertheless, if we try to apply this procedure to a Lagrangian with a 
topological mass term, we run into trouble. The equations obtained by 
using the Allen-Jacobson (A-J) basis do not make any sense. We can also 
see easily that the propagator for a vector with a topological mass term 
in 3 flat dimensions (which can be computed in a straightforward fashion), 
contains a term which clearly cannot be expressed in terms of the A-J basis.

What is lacking in the A-J basis is a term which contains contractions of 
the $\epsilon$ tensor with derivatives of the chordal distance. These 
contractions can only give a bitensor for $d=2p$ and for $d+1=2p$. This 
is consistent with the fact that we can only write a topological mass term 
for $p$ forms in these dimensions (as  $m A \wedge dA$ or as $m^2 A \wedge A$).

For $d=2p$ there are 2 types of equations of motion for $p$-forms which we can 
try to solve. The first one (also known as ``self duality in odd 
dimensions'' \cite{vanN}) is:
\be
{m \over p!}  \epsilon_{\mu_1\mu_2 ... \mu_p}^{\ \ \ \ \ \ \ \ \ 
\mu_{p+1} ... \mu_{d+1} }
\p_{\mu_{p+1}}A_{\mu_{p+2}...\mu_{d+1}}= m^2 A_{\mu_1...\mu_{p}} -  
J_{\mu_1 ... \mu_p} \label{1}
\ee
where $J_{\mu_1\dots \mu_p}$ is a covariantly conserved current 
which couples to the $p$-form.

The second one, comes from adding a topological mass term in a Maxwell 
action, and it describes gauge invariant forms:
\be
{1\over p!}D^{\lambda} D_{[\lambda}  A_{\mu_1 ... \mu_p]} = {m}  
\epsilon_{\mu_1\mu_2 ... \mu_p}^{\ \ \ \ \ \ \ \ \ \mu_{p+1} ... \mu_{d+1} }
D_{\mu_{p+1}}A_{\mu_{p+2}...\mu_{d+1}} - J_{\mu_1 ... \mu_p} \label{2}
\ee

For $d+1=2p$ it is possible to add to a Lagrangian, besides the normal 
mass term $m^2 A^2$, a term of the form $\tilde m ^2 A \wedge A$. While 
this is an interesting possibility, it does not seem to arise in physical 
situations (like compactifications of supergravity), and so we will not 
explore it completely here. Nevertheless we will comment in section 2.3
on the effect of such a term on the propagator. The complete 
investigation should be straightforward with the methods we have.

\subsection{Self duality in odd dimensions}
$${}$$

In Euclidean $AdS_{d+1}$ with the metric 
\be 
ds^2={1 \over z_0^2}(dz_0^2+\Sigma_{i=1}^d{dz_i^2} )\label{3}
\ee
invariant functions and tensors are most easily expressed in terms of the 
chordal distance
\be
u \equiv {(z_0-w_0)^2+(z_i-w_i)^2 \over 2 z_0 w_0}\label{4}
\ee
and derivatives thereof.

As explained in \cite{vanN}, when $p$ is odd, (\ref{1}) can be interpreted as 
the square root of the equation of motion of a real massive form field, in 
Minkowski signature.

When $p$ is even, in order for (\ref{1}) to be interpreted as the 
square root of an equation of motion with positive mass, a factor of $i$ 
has to be added to its left hand side. Equation (\ref{1}) is now complex, 
and describes a complex field. As explained in \cite{vanN}, this description 
is equivalent to that of a real field satisfying the massive Proca equation, 
and  therefore redundant. We will discuss however at the end of this section 
the propagator for this case.

Since we use Euclidean $AdS$, we need to analytically continue the 
equation of motion. Since we want the $\epsilon$ symbol not to change, we 
have to change the equation of motion, multiplying by $i$ wherever 
$\epsilon$ appears. The new equations describe a complex field, but they
should describe a real 
field when continued back into Minkowski space.

Thus, the equation satisfied by the propagator is: 
\bea
i m \epsilon_{\mu_1\mu_2 ... \mu_p}^{\ \ \ \ \ \ \ \ \ \mu_{p+1}... 
\mu_{d+1} } D_{\mu_{p+1}}G_{\mu_{p+2}...\mu_{d+1}; \mu'_1 ... \mu'_p}&=& 
m^2 G_{\mu_1...\mu_{p}; \mu'_1 ... \mu'_p} \nonumber\\&-& \delta(z,w)
(g_{[\mu_1 \mu'_1} g_{\mu_2 \mu'_2}  ... g_{\mu_p] \mu'_p})
\label{5}
\eea
where the square brackets in the source term denote antisymmetrization of 
unprimed indices, and $m$ is a dimensionless parameter.

Since the propagator is a maximally symmetric bitensor, conventional wisdom 
is to express it in terms of 2 antisymmetric bitensors:
\be
T^1_{\mu_1 ... \mu_p; \mu'_1 ... \mu'_p } = \p_{\mu_1}\p_{\mu'_{1}}u ...  
\p_{\mu_p}\p_{\mu'_{p}}u +\hbox{antipermutations of primed indices}\label{6}
\ee
\be
T^2_{\mu_1 ... \mu_p; \mu'_1 ... \mu'_p } = {1\over (p-1)!}(\p_{\mu_1}u 
\ \p_{\mu'_{1}}u \ \p_{\mu_2}\p_{\mu'_{2}}u ...  \p_{\mu_p}\p_{\mu'_p}u + 
\hbox{antipermutations of all indices})\label{7}
\ee  
Nevertheless, if we try to plug a combination of these two bitensors into 
(\ref{5}) 
we obtain nonsense. We realized however that $G$ could contain another $p-p$ 
bitensor, which can only exist for $d=2p$:
\be
T^3_{\mu_1 ... \mu_p;\mu'_1 ... \mu'_p } =  \epsilon_{\mu_1\mu_2 ... 
\mu_p}^{\ \ \ \ \ \ \ \ \ \mu_{p+1} ... \mu_{d+1} }  \p_{\mu_{p+1}}
\p_{\mu'_{1}}u \ ... \p_{\mu_{d}}\p_{\mu'_{p}}u \ \p_{\mu_{d+1}}u\label{8}
\ee

One may ask if there is yet another bitensor, obtained from $T^3$ by 
switching primed with unprimed indices. However, at a closer investigation 
this bitensor turns out to be proportional 
to $T^3$.

Thus, the propagator can be expanded as:
\be
G_{\mu_1 ... \mu_p;\mu'_1 ... \mu'_p }(u)=T^1[G(u)+p H(u)]+T^2 H'(u) + 
T^3 K(u)\label{9}
\ee
where the splitting of the coefficient of $T^1$ has been made knowing in 
advance that $p H(u)T^1+H'(u)T^2 $ can be expressed as an antisymmetrized 
covariant derivative acting on a (p-1,p) bitensor, and thus it drops out 
when the kinetic operator is applied on it.
For $z\neq w$, the Euclidean continuation of the equation for the propagator 
can be expressed as $* d G = - i m G$. We are using $G$ as both shorthand 
notation for the propagator bitensor, and as a scalar function of $u$. 
We can easily work out the actions of $d$ and $*$ on the terms of the 
propagator, using the formulas in the Appendix. Thus:
\bea
&&*d T^1 G = (-1)^p T^3 G' \label{10a}\\
&& d (T^1 p H + T^2 H') = 0 \label{10b}\\
&& *d (T^3 K) = T^1 [u(2+u)K' + (d-p+1)(1+u) K] - T^2 [(1+u) K' + 
(d-p+1)K] \nonumber\\ \label{10c}
\eea
The equation for the propagator implies:
\bea
&&(-1)^p G'= (-im) K \label{11a}\\
&& u(2+u) K'+(d-p+1)(1+u) K = (-im)(G+p H) \label{11b}\\
&& -K' (1+u) - (d-p+1) K = (-im) H' \label{11c}
\eea
Equation (\ref{11b}) also contains a source term coming from 
the $\delta$ function 
term in (\ref{5}). Combining (\ref{11a}) and (\ref{11c}) we 
obtain a relation between $G$ and 
$H$, which we can integrate once (fixing the integration constant so that 
both go to $0$ as $u \rightarrow \infty$) to give:
\be
(-1)^p m^2 H = G'(1+u)+(d-p)G\label{12}
\ee
From (\ref{12}), (\ref{11a}) and (\ref{11b}) we can obtain:
\be
u(u+2) G'' + (d+1) (1+u) G' + [p(d-p) + (-1)^p m^2]G = 0\label{13}
\ee
We can observe that for the cases when ``self duality in odd dimensions'' 
holds (odd $p$), $G$ and $H$ obey the same equations as in the case of 
the massive form propagator \cite{io2,asad}. 
This is not so 
surprising; after all, our equation of motion was the square root 
of the equation of motion for massive forms. Moreover, the source term in 
(\ref{13}) coming from (\ref{5}) is identical to the 
source term for the massive  
propagator. 
Thus, $G$ and $H$ will be the same as for massive forms in $d=2p$:
\be
G = {(-1)^p \G({d-1 \over 2})\over 4 \pi^{d+1\over2} u^{d-1\over2} } 
\ {C \ {}_2F_1 (m+1/2,m+1-p,2m+1,{2\over 2+u}) \over (2+u)^{m+1/2}}
\label{14}
\ee
where $C$ is a constant that normalizes the second fraction to 1 as $u 
\rightarrow 0$\footnote
{We note that for cases of interest to the AdS/CFT 
correspondence the hypergeometric function simplifies to 1, since $m=1$ 
for $AdS_5$
and $m=2$ for $AdS_7$.} (for the curious, 
$C= 2^{m}\Gamma(m+1)\Gamma(p-1/2)/[\sqrt{2\pi} \Gamma(m+p)]$.) 
$H$ and $K$ are given by (\ref {12}), and respectively (\ref{11a}). 
It is interesting to notice that $K$ satisfies the equation (\ref{13}) 
for $p+1$ 
forms in $2p+3$, so it is - like G - a hypergeometric function. 
Nevertheless, its normalization is given by (\ref{11a}).

When $p$ is even, the extra $i$ in (\ref{1}) makes the equation real in 
Euclidean coordinates. The factor $(-im)$ from (\ref{11a},\ref{11b},\ref{11c}) 
becomes 
$m$, which has the effect of changing $(-1)^p m^2$ to $(-1)^{p+1}m^2$ in 
equations (\ref{12}) and (\ref{13}). Since $p$ is now even, the last term of 
(\ref{13}) is the same as before, and (\ref{14}) is unchanged.


\subsection{Forms with a topological mass term}
$${}$$

Unlike  the previous case, when there was no gauge invariance, equation 
(\ref{2}) describes gauge invariant forms. As explained in \cite{fred}, 
working in 
the subspace of covariantly conserved currents makes gauge fixing unnecessary.
The equation for the propagator is obtained from (\ref{2}), remembering that 
Euclidean continuation introduces an $i$ multiplying the $\epsilon$ tensors:
\bea
&&{1\over p!}D^{\lambda} D_{[\lambda}  G_{\mu_1 ... \mu_p]; \mu'_1 ... \mu'_p} 
-i m \epsilon^{\mu_1\mu_2 ... \mu_{d+1} } D_{\mu_{p+1}}G_{\mu_{p+2}...
\mu_{d+1}; \mu'_1 ... \mu'_p } = \ \ \ \ \ \ \ \ \nonumber\\
&&\ \ \ \ \ \ \ \ \ \ \ \ \ \ =   - \delta(z,w) (g_{[\mu_1 \mu'_1} 
g_{\mu_2 \mu'_2}  ... g_{\mu_p] \mu'_p}) +  {1\over (p-1)!} 
D_{[\mu'_{1}}S_{\mu_{1}...\mu_{p}; \mu'_2 ... \mu'_p]}\label{15}
\eea
where the last term is a  diffeomorphism whose contribution vanishes when 
integrated against a covariantly conserved source.
Since $S$ can be expressed as: 
\be
S_{\mu_{1}...\mu_{p}; \mu'_2 ... \mu'_p} = S(u) [\p_{[\mu_1}\p_{\mu_2}
\p_{\mu'_2}u ... \p_{\mu_p]}\p_{\mu'_p}u]\label{16}
\ee
the last term can be written as $T^1 p S+T^2 S'$. For $z\neq w$, equation 
(\ref{15}) can be written in a more compact notation as:
\be
(-1)^p*d*d G - i m *d G = T^1 p S+T^2 S' \label{17}
\ee
We are expanding $G$ as in (\ref{9}), remembering that the terms containing 
$H$ can be written as the total divergence, and thus they are gauge 
artifacts.  Using the equations (\ref{10a}, \ref{10b}, \ref{10c}) we obtain 
after a few straightforward steps:
\bea
&&u(u+2)K''+(d+3)(1+u)K'+(1+p)(d-p+1)K=(-1)^p(im)G'   
\label{18a}\\
&&u(u+2)G''+(d-p+1)(1+u)G'= im[ u(u+2)K'+(d-p+1)(1+u)K] +p S  \nonumber\\
\label{18b}\\
&&(1+u)G''+(d-p+1)G'= im[(1+u)K'+(d-p+1)K] - S' \label{18c}  
\eea
As before, equation (\ref{18b}) contains a source term coming from 
the right 
hand side of (\ref{15}). In order to find $G$ and $K$ we have to do some 
manipulations  on the system (\ref{18a}, \ref{18b}, \ref{18c}). 
We first define $F$, such that 
$F' \equiv K $. Having done this we can integrate (\ref{18a}) and (\ref{18c}) 
once, setting the integration constants so that everything goes to $0$ as $u 
\rightarrow \infty$. We obtain
\be
u(u+2)F''+(d+1)(1+u)F'+p(d-p)F=(-1)^p(im)G\label{19} 
\ee
\be
(1+u)G'+(d-p)G= im[(1+u)F'+(d-p)F] - S \label{20} 
\ee
We combine (\ref{18b}) with (\ref{20}), and obtain:
\be
u(u+2)G''+(d+1)(1+u)G'+p(d-p)G=im[u(u+2)F''+(d+1)(1+u)F'+p(d-p)F ]
\label{21} 
\ee
which by using (\ref{19}) gives 
\be
u(u+2)G''+(d+1)(1+u)G'+[p(d-p)+(-1)^p m^2]G=0 \label{22}
\ee
Both (\ref{21}) and (\ref{22}) contain a source term at the right hand 
side. For odd 
$p$, (\ref{22}) is the massive form propagator equation, and $G$ is given by 
(\ref{14}). 
We will discuss the even $p$ situation later.
Equation (\ref{21}) implies that $G-imF$ satisfies the {\it{massless}} 
propagator equation, with the same source term, so $F$ will be proportional 
to the difference between the massive and the massless propagator. 
This makes sense; as $m\rightarrow 0$ we expect the propagator to 
approach the massless one, which does not contain $K$. Thus:
\be
K={-i \over m}(G'-G_{m=0}')\label{23}
\ee
where $G_{m=0}$ is given by setting $m=0$ in (\ref{14}).

As promised before, we now investigate the situation of even $p$. 
Naively, our theory can be defined for any $p$. However, 
if we square the equation satisfied by the 
field, $(-1)^p*d*dA=im *d A$ and call $B \equiv *dA$, we obtain $(-1)^p*d*dB = 
(-1)^{p+1} m^2 B$ which describes a field of positive mass only in odd 
dimensions. 
One way to remedy this is to modify (\ref{15}) by making the mass complex 
($m\rightarrow im$). This equation now describes a {\it complex} 
Minkowski field
of positive mass.

\subsection{Odds and ends}
$${}$$

There are two questions which we have not yet addressed. 
The first one has to do 
with the effect of the extra term in the bitensor ansatz on the other 
propagators computed so far. We can investigate the massive propagator, 
from which the massless one can be obtained as a limit.
We can plug the ansatz (\ref{9}) into the equation for the propagator:
\be
(-1)^p*d*dG=m^2 G \label{24}
\ee
and obtain the equations for $G$ and $H$, in the normal 
fashion. The equation for $K$ is decoupled from the equations for $G$ 
and $H$. $K$ satisfies the same equation as $G$, but {\it without} a source 
term. Therefore $K$ is zero. In general, when the equation of motion contains 
an even number of $*$'s(or $\epsilon$ symbols) the equation for $K$ decouples 
from the other equations, and $K$ vanishes because of the 
lack of a source term.

The second issue to address is the case $d+1=2 p$. Adding to the massive 
form Lagrangian a term of the form $\tilde m ^2 A \wedge A$ is legitimate. 
The equation of motion becomes:
\be
(-1)^p *d*dA = m^2 A + \tilde m^2 *A\label{25}
\ee
and the new ansatz for the propagator is:
\be
G_{\mu_1 ... \mu_p;\mu'_1 ... \mu'_p }(u)=T^1[G(u)+p H(u)]+T^2 H'(u) + 
T^4 K(u)
\label{26}
\ee
where
\be
T^4_{\mu_1 ... \mu_p;\mu'_1 ... \mu'_p} =  
\epsilon_{\mu_1\mu_2 ... \mu_p}^{\ \ \ \ \ \ \ \ \ \mu_{p+1} ... \mu_{d+1} }  
\p_{\mu_{p+1}}\p_{\mu'_{1}}u \ ... \p_{\mu_{d+1}} \p_{\mu'_{p}}u\label{27}
\ee

For $\tilde m \neq 0$ we obtain a set of coupled 
differential equations involving $G, H$ and $K$.
We do not solve this case here, since the presence of an $\tilde m$ term is 
more a logical possibility than something which arises naturals in a 
physical theory.

Also, this extra term has no effect on the other propagators computed so far.
When $\tilde m = 0$ the equation for $K$ 
decouples from the equations for $G$ and $H$, and thus  $K=0$, in absence 
of a source term. 
\section{Correlators}
$${}$$

We begin this section dedicated to computing correlators on 
$AdS_7$ involving the self-dual 3-form $S_{\mu\nu\rho}$ with a brief
review of the correspondence between the fields of supergravity and the 
$CFT$ operators. The fields of maximally $7d$ gauged ($SO(5)_g$) 
sugra couple on the boundary (via the $AdS-CFT$ correspondence) to 
the operators of the $6d$ (2,0) $CFT$. It is well known that the $6d$ (2,0) 
$CFT$ with
gauge group $SU(N)$ has 
no known lagrangian formulation, but the abelian version corresponds to 
the tensor
multiplet of (2,0) supersymmetry. The $CFT$ operators (characterized by their
Lorentz ($Spin(6)$) and by their R-symmetry $Sp(2)_R\sim SO(5)$ 
quantum numbers) 
are built out of the primary gauged invariant 
operators: $\phi^A$ (a vector under the R-symmetry group), $\psi$ 
(a spinor under the R-symmetry), $H_{ijk}$ (a singlet under the R-symmetry). 
These operators, which are all in the 
adjoint representation of $SU(N)$ transform under supersymmetry as a $(2,0)$
tensor multiplet \cite{lr}. 
Below we list the supergravity fields and the $CFT$ operators to 
which they couple:
\begin{itemize}
\item the scalars 
${\Pi^i}_A$ in the coset $Sl(5,R)/SO(5)_g$ (with the group $Sl(5,R)$ broken
after gauging to $SO(5)_c$) and in the $\bf 14$ of $SO(5)_g$ $\rightarrow$
$tr(\phi\phi)$
\item the spin 1/2 fields $\lambda_i$ in the ${\bf 16}$ of 
$SO(5)_g$ $\rightarrow$
$tr(\psi\phi)$
\item the self-dual 3-form $S_{\mu\nu\rho}^A$, in the ${\bf 5}$ of $SO(5)_g$
$\rightarrow$ $tr(H_{ijk}\phi^A)\equiv{\cal O}^A_{ijk}$
\item the gauge fields $B_\mu^{AB}$ in the adjoint ${\bf 10}$ of $SO(5)_g$
couple to the $R$-current $\rightarrow$ $tr(\bar\psi\gamma_i\psi)
\equiv J_i$
\item the four gravitini $\psi_\mu$ in the spinor ${\bf 4}$ representation 
of $SO(5)_g$ $\rightarrow$ $tr(\gamma^{jk}\psi H_{ijk})$
\item the graviton $g_{\mu\nu}$, which is  singlet under $SO(5)_g$, couples to 
the stress-energy tensor $\rightarrow$  
$tr(H_{mjk} H_{njk})+\dots$
\end{itemize}
\newpage
\subsection{2-point function}
$${}$$

In the first section we derived the propagator for a self-dual form in 
odd dimensions. In particular, in 7 dimensions we have:
\be
G(u)=c\cdot u^{-5/2}(2+u)^{-m-1/2}{}_2F_{1}(m+1/2,m-2;2m+1;2/(2+u))
\ee
with $c$ a constant. From the bulk propagator one can derive easily the 
bulk-to-boundary propagator: 
\be
S_{\mu_1\mu_2\mu_3}(w^0,{\bf w})=\lim_{\epsilon\rightarrow 0}\int_{M_\epsilon} 
d^6z\sqrt{h(z)}G_{\mu_1\mu_2\mu_3;\mu'_1\mu'_2\mu'_3}(w,z)(z^0)^{-m}
s^{\mu'_1\mu'_2\mu'_3}({\bf z})\label{s}
\ee
where $d^6 z \sqrt{h(z)} $ is the invariant volume element on the 
hypersurface $M_\epsilon=\{z^0=\epsilon\}$ and G(w,z) is the bulk-to-bulk
propagator. The field $S_{\mu_1\mu_2\mu_3}(w)$ 
as defined by (\ref{s}) satisfies its field equation for any finite $s_{\mu_1
\mu_2\mu_3}({\bf z})$. Moreover, the only nonvanishing components of the 
$s({\bf z})_{\mu_1\mu_2\mu_3}$ are the ones with indices ``on the boundary'' 
$\{i, j,\dots\}$.
In the limit $\epsilon\rightarrow 0$ we have $u\rightarrow \infty$, $G(u)
\rightarrow c u^{-m-3}$, $H(u)\rightarrow -(c/m) u^{-m-3}$ and $K(u)\rightarrow
(i(m+3)c/m) u^{-m-4}$.   
So,
\be
S_{ijk}(w)=\lim_{\epsilon\rightarrow 0}\int_{M_\epsilon} d^6z
G_{ijk;i'j'k'}(w,{\bf z}) s_{i'j'k'}({\bf z})\label{sijk}
\ee
where the metric ``on the boundary'' (used here for raising and lowering 
the primed 
indices) is flat, $g_{i'j'}({\bf z})=\eta_{i'j'}$, while the bulk metric is 
$AdS$, $g_{\mu_1\mu_2}(z)=(1/{z^0})^2 \eta_{\mu_1\mu_2}$ and $G(w,{\bf z})$ is 
the bulk-to-boundary propagator.
Substituting the propagator expression into (\ref{sijk}), and taking 
into account the various limits we get:
\bea
S_{ijk}(w)&=&c\int d^6z
\left[6\frac{m+3}{m}s^{3+m}2^{m+3} \frac{{w^0}^m}{[\epsilon^2+
({\bf z -w})^2)]^{3+m}}s_{ijk}({\bf z})\right.\nonumber\\
&+&\frac{i(3+m)}{2m}2^{m+4}\frac{{w^0}^{m+2}}{[{w^0}^2+({\bf z-w})^2]^{m+4}}
\epsilon_{ijki'j'k'}s_{i'j'k'}({\bf z})\nonumber\\
&-&\frac{i(3+m)}{m}2^{m+3}\frac{{w^0}^m}{[{w^0}^2+({\bf z-w})^2]^{m+3}}
\epsilon_{ijki'j'k'}s_{i'j'k'}({\bf z})\nonumber\\
&-&18 \frac{m+3}{m}({\bf w-z})_i \cdot ({\bf w-z})_{i'}2^{m+4}\frac{{w^0}^m}{
[{w^0}^2+({\bf z-w})^2]^{m+4}}s_{i'jk}({\bf z})\nonumber\\
&+&\left.3\frac{i(3+m)}{m}({\bf w-z})_{i'}({\bf w-z})_l 2^{m+4}\frac{{w^0}^m}{
[{w^0}^2+({\bf z-w})^2]^{m+4}}\epsilon_{ijkj'k'l}s_{i'j'k'}({\bf z})\right]
\nonumber\\{}\label{intermediar}
\eea
Use now the Schouten identity to rewrite the last term as
\bea
\epsilon_{ijki'j'k'}s_{li'j'}({\bf z})({\bf w-z})_l ({\bf w-z})_{k'}&=&
\epsilon_{ljki'j'k'}s_{li'j'}({\bf w-z})_i ({\bf w-z})_{k'}\nonumber\\
&+&\frac{1}{3}\epsilon_{ijki'j'l}s_{i'j'l} ({\bf w-z})^2
\eea
and by rearranging the terms of (\ref{intermediar}) we finally get
\bea
S_{ijk}(w)&=&6c\int d^6z \left(\frac{3+m}{m}\frac{2^{3+m}{w^0}^m}{
[{w^0}^2+({\bf z-w})^2]^{m+3}}\right)\nonumber\\
&&\left[\left(s_{ijk}+\frac{i}{3!}\epsilon_{ijki'j'k'}s_{i'j'k'}\right)-6
\frac{({\bf w-z})_i({\bf w-z})_{i'}}{[{w^0}^2+({\bf z-w})^2]}\left(
s_{i'jk}+\frac{i}{3!}\epsilon_{i'jklmn}s_{lmn}\right)\right]
\nonumber\\{}\label{sijkfinal}
\eea

The value which $S_{ijk}(w)$ takes on the boundary is obtained from 
(\ref{sijkfinal}) with $w^0\rightarrow 0$. 
Using further the limits (\ref{unu},\ref{patru})
we get the on-shell boundary value of the 7d self-dual tensor $S_{ijk}
({\bf w})$:
\be
\lim_{\epsilon\rightarrow 0} \epsilon^m S_{ijk}(\epsilon,{\bf w})= 
6c \pi^3\frac{2^{m+3}}{m(m+1)(m+2)}\left( s_{ijk}({\bf w})+\frac{i}{3!}
\epsilon_{ijki'j'k'}s_{i'j'k'}({\bf w})\right)\label{boundlim}
\ee 
We are not surprised to see that
on the boundary $S_{ijk}({\bf w})$ is self-dual (after all, we knew that
$S_{ijk}({\bf w})$ is source 
for a self-dual $CFT$ operator). It is interesting to note however, that 
even if we didn't impose any restriction on $s_{ijk}({\bf z})$, 
the form of the 
bulk-to-bulk propagator determined that only the self-dual part of 
$s_{ijk}({\bf z})$ propagates in the bulk. Alternatively, this can be seen
in the derivation of the bulk-to-boundary propagator in the footnote on page 
11, as a
constraint which relates the anti-self-dual part to the self-dual part.

The other on-shell components of $S_{\mu_1\mu_2\mu_3}(w)$, namely 
$S_{0ij}(w)$, can be determined similarly, and we obtain the following
bulk-to-boundary propagator:
\be
G_{\mu \nu\rho;lmn}(w;{\bf z})^{A;B}=c_1 \left[\frac{w^0}{(w-{\bf z})^2}
\right]^m
\partial_{[\mu}\frac{(w-{\bf z})_{i'}}{[(w-{\bf z})^2]^3}\delta^{i'|\nu\rho]}
_{lmn}\delta^A_B
\ee
where the Kronecker delta symbols are antisymmetrized with strength one, and
the indices $lmn$ are in the self-dual representation of the Lorentz group 
$SO(6)$.
We normalize the bulk-to-boundary  propagator such that for $w^0\rightarrow 0$,
$S_{ijk}\cdot (w^0)^m\rightarrow s_{ijk}^+({\bf w})$, with $s _{ijk}^+({\bf w})=
\frac{1}{2}(s_{ijk}({\bf w})+\frac{i}{3!}
\epsilon_{ijki'j'k'}s_{i'j'k'}({\bf w}))$ self-dual. 
Thus $c_1=(m+1)(m+2)(m+3)/\pi^3$.
Furthermore, using a notation which was first introduced in a paper 
by Freedman et al.
\cite{fmmr}, we can rewrite the bulk-to-boundary propagator in a 
manifestly conformal-covariant form:
\be
G_{\mu \nu\rho ;lmn}(w;{\bf z})^{AB}=c_1\left[\frac{w^0}{(w-{\bf z})^2}
\right]^m
\frac{J_{\mu l}(w-{\bf z})}{(w-{\bf z})^2}
\frac{J_{\nu m}(w-{\bf z})}{(w-{\bf z})^2}
\frac{J_{\rho n}(w-{\bf z})}{(w-{\bf z})^2}\delta^A_B
\ee
where the symmetry in the indices $\mu \nu\rho$ and $lmn$ is the same on both 
sides. The tensor $J_{\mu\nu}(w)$ is related to the inversion $w^\mu\rightarrow
(1/w'^2) {w'}^{\mu}$ Jacobian
\be
\frac{\partial w'^\mu}{\partial w^\nu}=w'^2 J_{\mu\nu}(w')
\ee

Clearly, for a field obeying a first order differential equation, the bulk 
action is vanishing on-shell. This is a situation which was first 
encountered for spin 1/2 fields \cite{hs,mv}. The resolution which was 
proposed for obtaining non-zero 2-point functions (in accord with the $CFT$ 
calculations) was to supplement the bulk action with a boundary term 
which does not break the $AdS_{d+1}$ isometry group $O(d+1,1)$. 
This proposal was justified afterwards \cite{af1,hen} 
by imposing the variational 
principle on a manifold with boundary. On the boundary, we need to specify
only the self-dual piece of $S_{ijk}^+$
\footnote{
We can understand the result (\ref{sijkfinal}) from a different perspective.
The (linearized) field equation in momentum space
$$
\left[(x^0)^2\partial_{0}^2 -x^0\partial_0 -m(m\pm 2)-(x^0)^2 {\bf k}^2\right]
S_{ijk}^{\pm}(x^0,{\bf k})=0
$$
has a unique solution (which falls off at infinity) written in terms of the 
modified Bessel function 
$$
S^\pm_{ijk}(x^0,{\bf k})=\frac{x^0 K_{m\pm 1}(x^0 k)}{\epsilon 
K_{m\pm 1}(\epsilon k)}s^\pm_{ijk}({\bf k})
$$
However, as observed in \cite{af2}, there is a constraint (obtained from
another field equation)
which relates the 
self-dual to the anti-self-dual boundary values. In $7d$, the constraint yields
$$ 
s_{ijk}^-({\bf k})=-\frac{K_{m-1}(\epsilon k)}{K_{m+1}(\epsilon k)}\left(
s_{ijk}^+({\bf k})-\frac{6k_{i'}k_{[i} s_{jk]i'}^+({\bf k})}{k^2}\right)
$$
Note that $s_{ijk}^-$ vanishes if $s_{ijk}^+$ is kept
finite for $\epsilon\rightarrow 0$.

We finally get that on shell
$$
S_{ijk}(x^0,{\bf k})\propto k^m \left(m K_m(x^0 k)s_{ijk}^+({\bf k})+
\frac{3 x^0k_{i'}k_{[i}s_{jk]i'}^+({\bf k})K_{m-1}(x^0 k)}{k}\right)
$$
which is nothing else but the Fourier transform of (\ref{sijkfinal}).
}
, and so, the anti-self-dual
part is free to vary off-shell.
Therefore, in order to cancel the remaining boundary term $-m/2\oint S^+\delta
S^-$ in the variation of the action, we will add the following boundary 
term to the $7d$ gauged sugra action:
\be
{\cal S}_1=\frac{m}{4}\lim_{\epsilon\rightarrow 0}\int_{M_\epsilon} d^6z 
\sqrt{h(z)}
S_{\mu_1\mu_2\mu_3}(z)S^{\mu_1\mu_2\mu_3}(z)\label{bound}
\ee
where, as before, $d^6z \sqrt{h(z)}$ is the invariant volume element on
the hypersurface $M_\epsilon=\{z^0=\epsilon\}$ infinitesimally close to the
boundary of the $AdS$ space. 

On-shell, ${\cal S}_1$ becomes:
\bea
&&{\cal S}_1=\frac{m}{4} c_1\lim_{\epsilon\rightarrow 0}\int_{M_\epsilon} 
d^6z S_{ijk}(z)\epsilon^m\int d^6w \frac{1}{(\epsilon^2
+{\bf w}^2)^{m+3}}
\nonumber\\ 
&&\left(s_{ijk}^+(w)-6
\frac{({\bf w-z})_i({\bf w-z})_{i'}}{[\epsilon^2+({\bf z-w})^2]}
s_{i'jk}^+(w)\right)
\label{s1}
\eea
where we used that 
$\lim_{\epsilon\rightarrow 0} \epsilon^m S_{\mu_1\mu_2\mu_3}
(z)$ is nonvanishing only for the $S_{ijk}(z)$ components. The limit was
already evaluated in (\ref{boundlim}), and so we get:
\bea
{\cal S}_1&=&\frac{\pi^3 c_1^2 m}{4(m+1)(m+2)(m+3)}\int d^6z \int d^6w 
s_{ijk}^+({\bf z})\frac{1}
{(({\bf z-w})^2)^{m+3}}\left(s_{ijk}^+({\bf w})\right.\nonumber\\
&&\left.-6\frac{({\bf w-z})_i
({\bf w-z})_{i'}}{({\bf z-w})^2} s_{i'jk}^+({\bf w})\right)
\nonumber\\{}\label{s1f}
\eea

The generating functional of the $CFT$ operators is equal, by the $AdS-CFT$ 
correspondence, to the supergravity partition function which in the 
classical limit is
\be
Z_{AdS_7}[s_{ijk}]=\exp(-{\cal S}_1-{\cal S}_{7d\; sugra})
\ee
Thus, the 2-point function of the self-dual $CFT$ operator ${\cal O}^A_{ijk}=
tr(\phi^A H_{ijk})$ is
\bea
&&\langle {\cal O}^A_{ijk}({\bf z}) {\cal O}^B_{lmn}({\bf w})\rangle
\nonumber\\
&&=\frac{\pi^3 c_1^2 m}{2(m+1)(m+2)(m+3)}\delta^A_B
\frac{1}{(({\bf z-w})^2)^{m+3}}\left(\delta^{ijk}_{lmn}-6\frac{({\bf w-z})^{[i}
({\bf w-z})_{i'}}{({\bf z-w})^2}\delta^{i'jk]}_{lmn}\right)\nonumber\\
\label{2point}
\eea
where
the symmetry in the indices $ijk$ respectively $lmn$ is the same on
both sides of (\ref{2point}). 
This is the 2-point function for a CFT operator with scaling dimension 
\be
\Delta=p+m=3+m
\ee
in the self-dual tensorial representation of the (Euclidean) Lorentz group 
$O(6)$.
In general for a $CFT$ operator with scaling dimension $\Delta$ and in some 
arbitrary representation of the Lorentz group, conformal invariance fixes the 
2-point function to be \cite{hs,fggp}
\be
\langle O({\bf z}) O({\bf w})\rangle=\frac{1}{(({\bf z-w})^2)^\Delta}
R({\bf w, z})
\ee
where $R({\bf w, z})$ is the representation matrix of the $O(d)$ element
\be
R^i_j({\bf w, z})=\delta^i_j-\frac{2}{({\bf z-w})^2}({\bf w-z})^{i}
({\bf w-z})_{j}
\ee
\subsection
{3-point functions: $\langle {\cal O}_{ijk}{\cal O}_{lmn} J_p \rangle $
and $\langle {\cal O}_{ijk}J_mJ_n \rangle$}
$${}$$

For this section, the relevant part of the action of maximal 
$({\cal N}=4)$ gauged $7d$ sugra is\footnote{Euclidean signature:
$\epsilon\rightarrow i\epsilon$}:
\be
\int d^7w [\frac{im}{48} \epsilon^{\mu_1\dots\mu_7}S_{\mu_1\mu_2\mu_3}^A
F_{\mu_4\dots\mu_7}^A+\frac{1}{16\sqrt{3}}\epsilon^{\mu_1\dots\mu_7}
\epsilon_{ABCDE}S_{\mu_1\mu_2\mu_3}^A F_{\mu_4\mu_5}^{BC} F_{\mu_6\mu_7}^{DE}
+\dots]
\ee
where $F_{\mu_1\dots\mu_4}^A=4(D_{[\mu_1} S_{\mu_2\mu_3\mu_4]}^A+gB^{AB}_{[
\mu_1}S_{\mu_2\mu_3\mu_4]}^B)$ 
is the gauge covariant field strength of 
$S_{\mu_1\mu_2\mu_3}^A$, and $F_{\mu\nu}^{AB}$ is the gauge covariant field
strength of the gauge fields $B_\mu^{AB}$. The coupling constant is $g$ and 
it equals $2m$.

To compute the 3-point functions, inspired by \cite{fmmr},
we use a conformal-covariant bulk-to-boundary propagator for the gauge fields 
\be
G_{\mu i}^{AB;CD}(w,{\bf z})=c_2\left(\frac{w^0}{(w-{\bf z})^2}\right)^4
\frac{J_{\mu i}(w-{\bf z})}{(w-{\bf z})^2}\delta^{AB}_{CD}
\ee
with $c_2=\frac{\Gamma(d)}{2\pi^{d/2}\Gamma(d/2)}$.
The 3-point function $ \langle {\cal O}{\cal O} J \rangle$ on the $AdS$ side
of the correspondence reads:
\bea
\langle {\cal O}_{ijk}^A({\bf z_1}){\cal O}_{lmn}^B({\bf z_2}) 
J_p^{CD}({\bf z_3}) \rangle&=&\frac{igm{c_1}^2 c_2}{12}\int d^7w 
\epsilon^{\mu_1\dots\mu_7} G_{\mu_1\mu_2\mu_3;ijk}^{EA}(w;{\bf z_1})\nonumber\\
&&G_{\mu_4 p}^{EF;CD}(w;{\bf z_3}) G_{\mu_5\mu_6\mu_7;lmn}^{FB}(w;{\bf z_2})
\nonumber\\ &&+{\rm perm.} (ijk \leftrightarrow lmn, {\bf z_1}\leftrightarrow
{\bf z_2}, A\leftrightarrow B)
\label{3p1}
\eea 

Following \cite{fmmr} we will use conformal invariance to fix ${\bf z_3}=0$,
and we will make a change of variable by inverting the bulk and the boundary 
${\bf z_1 , z_2}$ points: $w^\mu\rightarrow (1/w'^2) w'^\mu $,
${\bf z_1}\rightarrow (1/{\bf z_1}^2) {\bf z_2}$,
${\bf z_2}\rightarrow (1/{\bf z_2}^2) {\bf z_3}$.
The inversion property which the tensors $J_{\mu\nu}$ satisfy will also be 
used:
\be
J_{\mu i}(w-{\bf z})=J_{\mu\rho}(w') J_{\rho j}(w'-{\bf z'}) J_{ ji}({\bf z'})
\label{inversion}
\ee
Finally, we will combine the factors of $J_{\mu_1 i}(w'),...$ with the ``flat
space''
epsilon symbol into $\det{[J(w')]}\epsilon_{ijk\nu_4\dots\nu_7}$.
The determinant can be easily evaluated (by induction, for example) to be -1.
Substituting the various propagators into (\ref{3p1}), and jumping the 
intermediary steps described above, the 
integral in the 3-point function becomes:
\bea
&&
-\frac{igm{c_1}^2 c_2}{12}\int {d^7w'}\delta^{AB}_{CD}
\epsilon_{\nu_1\nu_2\nu_3 p\nu_5\dots\nu_7} (w'^0)^{2m+4}
\frac{{\bf z'_1}^{2(m+3)}}{(w'-{\bf z'_1})^{2(m+3)}}
\frac{{\bf z'_2}^{2(m+3)}}{(w'-{\bf z'_2})^{2(m+3)}}
\nonumber\\
&& J_{\nu_1 i'}(w'-{\bf z'_1}) J_{ii'}({\bf z'_1}) 
   J_{\nu_2 j'}(w'-{\bf z'_1}) J_{jj'}({\bf z'_1})
   J_{\nu_3 k'}(w'-{\bf z'_1}) J_{kk'}({\bf z'_1})\nonumber\\
&& J_{\nu_5 l'}(w'-{\bf z'_2}) J_{ll'}({\bf z'_2}) 
   J_{\nu_6 m'}(w'-{\bf z'_2}) J_{mm'}({\bf z'_2})
   J_{\nu_7 n'}(w'-{\bf z'_2}) J_{nn'}({\bf z'_2})\nonumber\\
&&   +{\rm perm.}
\label{3p2}
\eea
where the factors of $w'$ canceled as expected.
Furthermore, we shift the integration variable such that all dependence 
on ${\bf z'_1}$ and ${\bf z'_2}$ appears through ${\bf z'_1}-
{\bf z'_2}$, and we use the integrals listed in Appendix 3. In the last step 
we restore the ${\bf z_3}$ dependence using the translational invariance of  
the 3-point function: ${\bf z'_1}\rightarrow {\bf z'_{13}}\equiv
{\bf (z_1-z_3)'}$,
${\bf z'_1}-{\bf z'_2} \rightarrow {\bf t}\equiv {\bf (z_1-z_3)'}-
{\bf (z_2-z_3)'}$, etc.
Thus the  3-point function $\langle J_p^{CD}({\bf z_3}) 
{\cal O}_{ijk}^A ({\bf z_1}){\cal O}_{lmn}^B ({\bf z_2})\rangle$ is 
\bea
&&-\frac{igm{c_1}^2 c_2}{12}\frac{\pi^3}{10(m+3)^2 (m+2)(m+1)}\delta^{AB}_{CD} 
\epsilon_{pj'k'rm'n'}({z_{13}}{z_{23}})^{-2(m+3)}
\nonumber\\&&J_{ii'}({\bf z'_{13}}) J_{jj'}({\bf z_{13}}) 
J_{kk'}({\bf z_{13}}) J_{ll'}({\bf z_{23}})J_{mm'}({\bf z_{23}})
J_{nn'}({\bf z_{23}}){\bf t}^{2m}\nonumber\\
&&\left((\delta_{rl'}{\bf t_{i'}}+
\delta_{ri'}{\bf t_{l'}}) 2m -6(\delta_{rl'}{\bf t_{i'}}+
\delta_{ri'}{\bf t_{l'}}+\delta_{i'l'}{\bf t_{r}})
-6\frac{{\bf t_{i'}}{\bf t_{l'}}{\bf t_{r}}}{t^2}(2m+2)\right)\nonumber\\
\eea
This expression can be further simplified by using the self-duality in 
the indices 
$(ijk)$, $(lmn)$, and whenever the case, by rearranging indices with 
the Schouten identity. We obtain:
\bea
&&-\frac{gm{c_1}^2 c_2}{2}\frac{\pi^3}{10(m+3)^2 (m+2)(m+1)}\delta^{AB}_{CD} 
({ z_{13}}{ z_{23}})^{-2(m+3)}\nonumber\\
&&J_{jj'}({\bf z_{13}}) 
J_{kk'}({\bf z_{13}}) J_{mj'}({\bf z_{23}})
J_{nk'}({\bf z_{23}})t^{2m-2}\nonumber\\
&&\left[(
J_{ip}({\bf z_{13}}) J_{ll'}({\bf z_{23}}){\bf t_{l'}}+
J_{ii'}({\bf z_{13}})J_{lp}({\bf z_{23}}){\bf t_{i'}})\left((2m-2
\right.\right.\nonumber\\
&&\left.\left.-(2m+2)\right)
+\left(\frac{{\bf t_{i'} t_r t_p}}{t^2}J_{ii'}({\bf z_{13}})J_{lr}
({\bf z_{23}})\right.\right.\nonumber\\
&&\left.\left.+\frac{{\bf t_{l'} t_r t_p}}{t^2}J_{ir}({\bf z_{13}})J_{ll'}
({\bf z_{23}})\right)(2m+2)\right]
\eea

Let's now turn to the 3-point function $ \langle {\cal O} J J \rangle$. 
The $AdS$ correlator is given by
\bea
\langle {\cal O}^A_{ijk}({\bf z_1}) J_l^{BC}({\bf z_2}) J_m^{DE}({\bf
z_3})\rangle &=&\frac{1}{4\sqrt{3}}c_1{c_2}^2\epsilon^{FGHIJ}\int d^{d+1}
\epsilon^{\alpha\beta\gamma\delta\epsilon\eta\zeta} G_{\alpha\beta\gamma 
; ijk}^{FA}(w;{\bf z_1})\nonumber\\&&
\partial_{\delta}G_{\epsilon ;l}^{GH;BC}(w;{\bf z_2})
\partial_{\eta}G_{\zeta ;m}^{IJ;DE}(w;{\bf z_2})
\nonumber\\&& +{\rm perm.}(BC\leftrightarrow DE, l\leftrightarrow m, 
{\bf z_2}\leftrightarrow {\bf z_3})
\eea
In a manner entirely analogous to the previous calculation, we use conformal 
invariance to fix ${\bf z_1}=0$, and then invert bulk and boundary points.
We use the inversion relation (\ref{inversion}) and the fact that the curl of 
the gauge field propagator also transforms covariantly, according to 
\be
\partial_{[\mu }G_{\nu i]} (w, {\bf z})= w'^2J_{\mu\rho}(w')w'^2 J_{\nu\sigma}
(w')J_{ki}
({\bf z'}){\bf z'}^{2(d-1)}\partial '_{[\rho}G_{\sigma] k}(w',
{\bf z'})
\ee
and again combine all the resulting factors of $J_{\mu i}(w')$ together with 
the $\epsilon$ symbol into a determinant which gives $-1$. The factors of 
w' cancel again, and we are left with the 3-point function
\bea
&&-\frac{1}{4\sqrt{3}}c_1{c_2}^2 \epsilon_{ABCDE}\frac{1}{({\bf z_2}^2)^{(d-1
)}({\bf z_3}^2)^{(d-1)}}J_{ll'}({\bf z_2})J_{mm'}({\bf z_3})\nonumber\\&&
\int d^{d+1}w \epsilon^{ijk\delta '\epsilon '\eta '\zeta '}
w_0 '^m\partial '_{\delta '}\left(\frac{w_0'^{d-2}}{(w'-{\bf z_2'})^{2(d-1)}}
J_{\epsilon ' l'}(w', {\bf z_2'})\right)\nonumber\\&&
\partial '_{\eta '}\left(\frac{w_0'^{d-2}}{(w'-{\bf z_3'})^{2(d-1)}}
J_{\zeta ' m'}(w', {\bf z_3'})\right)+{\rm perm.}
\eea
We then rewrite the summation over $\delta '\epsilon '\eta '\zeta '$ as 
summation over $l'',m'',n''$ and $0$ and notice that we can replace the 
partial derivatives $\partial '_{\delta '}$ and $\partial '_{\eta '}$, 
w.r.t. $w'$ with partial derivatives w.r.t. ${\bf z_2}$ and ${\bf z_3}$ 
respectively, if $\delta ' $ and $\eta '$ are not zero. In the case 
neither is zero, both  derivatives get out of the integral and 
then the antisymmetry kills them (the integral is a function of ${\bf z_2}
-{\bf z_3}$, so the derivatives are the same). Thus only the case when one of 
the derivatives is w.r.t. $w_0'$ remains, and we get
\bea
&&-\frac{1}{4\sqrt{3}}c_1{c_2}^2 \epsilon_{ABCDE}\frac{1}{({\bf z_2}^2)^{(d-1
)}({\bf z_3}^2)^{(d-1)}}J_{ll'}({\bf z_2})J_{mm'}({\bf z_3})\nonumber\\&&
\epsilon^{ijk l''m''n''}\frac{\partial}{\partial y^{n''}}\int d^{d+1}w'
\frac{w_0'^{m+2d-3}}{(w'-{\bf z_2'})^{2(d-1)}(w'-{\bf z_3'})^{2(d-1)}}
J_{l'l''}(w',{\bf z_2})\nonumber\\&&
\left((d-2)w_0'^{-2}J_{m'm''}(w', {\bf z_3})-\frac{2(d-1)}{(w'-{\bf z_3'})^{2}}
J_{m'm''}(w', {\bf z_3}) \right.\nonumber\\&&\left.
+4\frac{(w'-{\bf z_3'})^{m'}(w'-{\bf z_3'})^{m''}}
{(w'-{\bf z_3'})^{4}}\right)
\eea
Further using the identity
\be
\frac{(w'-{\bf z'})^{i}(w'-{\bf z'})^{j}}{(w'-{\bf z'})^{2a}}
=\frac{1}{4(a-1)(a-2)}\frac{\partial}{\partial z'^i}\frac{\partial}{\partial
z'^j}\frac{1}{(w'-{\bf z'})^{2(a-2)}}+\frac{1}{2(a-1)}\frac{\delta_{ij}}{
(w'-{\bf z'})^{2(a-1)}}
\ee
and the fact that, by the same antisymmetry argument as above, 
all extra partial 
derivatives w.r.t. ${\bf z_2'}$ and ${\bf z_3'}$ vanish, we get
\bea
&&-\frac{1}{4\sqrt{3}}c_1{c_2}^2 \epsilon_{ABCDE}\frac{1}{({\bf z_2}^2)^{(d-1
)}({\bf z_3}^2)^{(d-1)}}J_{ll'}({\bf z_2})J_{mm'}({\bf z_3})\nonumber\\&&
\epsilon^{ijkl'm'n''}\frac{\partial}{\partial t^{n''}} t^{m-d}\frac{(d-2)^2}{
d-1}\left[\frac{d-2}{d-1}C(m+2d-5, d-1, d-1)\right.\nonumber\\&&\left.
-2C(m+2d-3, d-1, d)\right]
\eea
where the constant $C(a,b,c)$ is defined in (A.16).
In the last step we restore the ${\bf z_1}$ dependence in $<{\cal O}JJ>$
\bea
&&
\langle {\cal O}^A_{ijk}({\bf z_1}) J_l^{BC}({\bf z_2}) J_m^{DE}({\bf
z_3})\rangle=\nonumber\\
&&-\frac{1}{4\sqrt{3}}c_1{c_2}^2\frac{\pi^{d/2}}{2}\frac{m(d-2)^2}{(\Gamma (d))
^2}\Gamma (\frac{m+2d-4}{2})\Gamma(\frac{d-m+2}{2})\frac{(\Gamma(\frac{d+m-2}
{2}))^2}{\Gamma(d+m-2)}\nonumber\\&&
 \epsilon_{ABCDE}\frac{1}{(({\bf z_2}-{\bf z_1})^2)^{(d-1
)}(({\bf z_3}-{\bf z_1})^2)^{(d-1)}}J_{ll'}({\bf z_2}-{\bf z_1})J_{mm'}
({\bf z_3}-{\bf z_1})\nonumber\\&&
\epsilon^{ijkl'm'n''}\frac{{\bf t}^{n''}}{t^{m-d+2}}
\eea
where ${\bf t}=({\bf z_2}-{\bf z_1})/({\bf z_2}-{\bf z_1})^2-({\bf z_3}-
{\bf z_1})/({\bf z_3}-{\bf z_1})^2$.

However, our 3-point function computation was too naive,
since we didn't carefully impose that the sugra fields satisfy their field 
equations. By doing so, we will discover that the boundary term (\ref{bound})
will give contributions to both correlators $<{\cal OOJ}>$ and $<{\cal OJJ}>$.

To be able to appreciate this point, we will look at a simple example of a 
$\lambda \phi^3$ scalar field theory, (understanding that this will apply to 
the case of general fields)
\be
S=\int d^{d+1}x[ \frac{1}{2}
((\partial_{\mu}\phi)^2 +m^2\phi^2) +1/3\lambda\phi^3]
\ee
with equation of motion $(-\Box +m^2)\phi =\lambda\phi^2$. The solution 
which takes the boundary value $\phi_0({\bf y})$ is of the type
\bea
&&\phi(x)=\phi^{(0)}(x)+\phi^{(1)}(x)+...\nonumber\\
&& \phi^{(0)}(x)=\int d^d{\bf y} 
G(x,{\bf y})\phi_0({\bf y})
\eea
where $G(x,{\bf y})$ is the bulk-to-boundary propagator 
and $\phi^{(1)}$ is zero on the 
boundary. In the $AdS-CFT$ correspondence, the 3-point function will a priori
have a contribution not only from the cubic term, $\int \lambda(\phi^{(0)})^3$,
but also from the kinetic piece, because $\phi^{(1)}(x)=\lambda\int d^{d+1}y
G(x,y)(\phi^{(0)}(y))^2$ ($G(x,y)$ is the bulk-to-bulk propagator).
If we take 
into account the fact that $\phi^{(0)}$ satisfies the free equation of motion
the bulk piece in the kinetic term vanishes. Moreover, the boundary term 
generated by partial integrations
vanishes because  $\phi^{(1)}$ is zero on the boundary . 
Hence the only contribution to the 3-point function 
comes from the cubic term in the action. 
Since we didn't need the explicit form of $\phi^{(1)}$ 
for this argument, the same argument goes through for 
3-point couplings to other
fields (still no contribution from the $\phi$ kinetic term).

Let us carry the same analysis when the kinetic action is 
linear in derivatives. All the terms in the action 
quadratic in $\phi$ (including the interactions with other fields) don't 
contribute to the 3-point functions if we use the complete equation of motion 
(unlike  the previous case, there is no need for partial integration, so no 
boundary terms). In fact now we can use the complete equation of motion 
to replace the kinetic terms with interaction terms. So the net effect 
of the kinetic piece will be to modify the coefficients of the interaction 
terms. It would appear that in this case, all 3-point functions quadratic
in $\phi$ vanish! This is not so in general, since to make the 2-point 
function nonzero we need to add a (pseudo)boundary term to the action. There
are two ways that this term can contribute. Either it is of the type 
$\int d^{d+1}x \partial_{\mu}(\partial^{\mu}\phi ....)$ (real boundary term, 
but with an extra derivative on $\phi$, so that we get a nonzero 
contribution if we replace $\phi $ by $\phi^{(1)}$), or it is a pseudo-boundary
term of the type $\int d^{d+1}\delta(z_0-\epsilon)(...)$, defined on the 
``regulated'' boundary $z_0=\epsilon$. The latter happens in our case.

Consider the part of $7d$ gauged sugra action 
containing only the self-dual 3-form and the gauge fields:
\bea
&&\int d^7z\sqrt{G^{-1}}\left(-\frac{1}{4}(F_{\mu\nu}^{AB})^2
+\frac{m^2}{2}
(S_{\mu\nu\rho}^A)^2+
\frac{im}{48\sqrt{G^{-1}}}\epsilon^{\mu_1\dots\mu_7}
(S_{\mu_1\mu_2\mu_3}^A F_{\mu_4\dots\mu_7}^A
\right.\nonumber\\
&&\left.+\frac{\sqrt 3}{im}\epsilon_{ABCDE}
S_{\mu_1\mu_2\mu_3}^A F_{\mu_4\mu_5}^{BC}F_{\mu_6\mu_7}^{DE})\right)
+\frac{m}{4}\int_{M_\epsilon} d^6z \sqrt{h(z)}(S_{\mu\nu\rho}^A)^2
\label{7dact}
\eea

Imposing that $S_{\mu\nu\rho}$ satisfies its field equation (to linear 
order in the coupling constant, and to second order in fields, since we are
interested in 3-point functions) we obtain
\be
S_{\mu\nu\rho}=S^{(0)}_{\mu\nu\rho}+S^{(1)}_{\mu\nu\rho}
\ee
where $S^{(0)}$ satisfies the linearized field equation, while $S^{(1)}$ is
given by
\bea
&&{S^{(1)}}_{\mu\nu\rho}^A(z)=-\int d^7 w {G^{(0)}}_{\mu\nu\rho;\mu'
\nu'\rho'}^{AA'}(z,w)
\epsilon^{\mu'\nu\rho'\mu_4\dots\mu_7}
\frac{img}{6}{B^{(0)}}_{\mu_4}^{AB}(w) {S^{(0)}}_{\mu_5\mu_6\mu_7}^B (w)
\nonumber\\
&&-\int d^7 w \epsilon^{\mu'\nu'\rho'\mu_4\dots\mu_7} 
{G^{(0)}}_{\mu\nu\rho;\mu'\nu'\rho'}^{AA'}(z,w)
\frac{1}{16\sqrt 3}\epsilon^{A'BCDE} {F^{(0)}}_{\mu_4\mu_5}^{BC}(w)
{F^{(0)}}_{\mu_6\mu_7}^{DE}(w)\nonumber\\
\label{sfeq}
\eea
Substituting (\ref{sfeq}) into (\ref{7dact}) we see that the $SBB$ bulk term
gets modifies by a factor 1/2 and becomes
\be
\int d^7 w \frac{1}{32\sqrt 3}\epsilon^{ABCDE}\epsilon^{\mu_1\dots\mu_7}
{S^{(0)}}_{\mu_1\mu_2\mu_3}^A {F^{(0)}}_{\mu_4\mu_5}^{BC} 
{F^{(0)}}_{\mu_6\mu_7}^{DE}\label{sb1}
\ee
whereas the $SSB$ term gets killed, as we mentioned in the general discussion.
The kinetic term for the $B$'s doesn't contribute to the 3-point functions, 
because it has two derivatives, and the previous discussion applies.
Finally, the $S$ boundary term  yields a contribution to the 3-point 
functions from its 
$S^{(0)}S^{(1)}$ piece
\be
\frac{m}{2}\int_{M_\epsilon} d^6z \sqrt{h(z)} {S^{(0)}_{\mu\nu\rho}}^A
{S^{(1)\mu\nu\rho}}^A
\ee
which becomes a bulk integral after substituting (\ref{sfeq}) 
\bea
&&-\frac{m}{2}\int d^7 z \epsilon^{\mu_1\dots\mu_7} 
{S^{(0)}}_{\mu_1\mu_2\mu_3}^A (z)\left(
\frac{img}{6}{B^{(0)}}_{\mu_4}^{AB}(z) {S^{(0)}}_{\mu_5\mu_6\mu_7}^B (z)
\right.\nonumber\\&&+\left.
\frac{1}{16\sqrt 3}\epsilon^{ABCDE} {F^{(0)}}_{\mu_4\mu_5}^{BC}(z)
{F^{(0)}}_{\mu_6\mu_7}^{DE}(z)\right)\label{sb2}
\eea
We note here that it is crucial this was a pseudo-boundary term (defined
at $z_0=\epsilon$). If it was a real boundary term, one could have used that 
$S^{(1)}=0$ on the boundary to conclude that it doesn't contribute.
Adding (\ref{sb1}) to (\ref{sb2}) we get the corrected coefficients of 
the 3-point functions on the AdS side. Thus the (corrected) 3-point functions 
are:
\bea   
&&\langle J_p^{CD}({\bf z_3})
{\cal O}_{ijk}^A ({\bf z_1}){\cal O}_{lmn}^B ({\bf z_2})\rangle=\nonumber\\
&&2 gm^2{c_1}^2 c_2\frac{\pi^3}{10(m+3)^2 
(m+2)(m+1)}\delta^{AB}_{CD} ({ z_{13}}{ z_{23}})^{-2(m+3)}\nonumber\\
&&J_{jj'}({\bf z_{13}}) 
J_{kk'}({\bf z_{13}}) J_{mj'}({\bf z_{23}})
J_{nk'}({\bf z_{23}})t^{2m-2}\nonumber\\
&&\left[-(
J_{ip}({\bf z_{13}}) J_{ll'}({\bf z_{23}}){\bf t_{l'}}+
J_{ii'}({\bf z_{13}})J_{lp}({\bf z_{23}}){\bf t_{i'}})+
(m+1)\frac{{\bf t_{i'} t_r t_p}}{t^2}J_{ii'}({\bf z_{13}})J_{lr}
({\bf z_{23}})\right]\nonumber\\\label{3pjoo}\\
&&\langle {\cal O}^A_{ijk}({\bf z_1}) J_l^{BC}({\bf z_2}) J_m^{DE}({\bf
z_3})\rangle=\nonumber\\
&&-\frac{c_1{c_2}^2 (-m+1)\pi^3 m}{\sqrt{3} (5!)^2}
\Gamma (\frac{m+8}{2})\Gamma(\frac{4-m}{2})\frac{(\Gamma(\frac{4+m}
{2}))^2}{\Gamma(4+m)} \nonumber\\&&
 \epsilon_{ABCDE}\frac{1}{(({\bf z_2}-{\bf z_1})^2)^{(d-1
)}(({\bf z_3}-{\bf z_1})^2)^{(d-1)}}J_{ll'}({\bf z_2}-{\bf z_1})J_{mm'}
({\bf z_3}-{\bf z_1})\nonumber\\&&
\epsilon^{ijkl'm'n''}\frac{{\bf t}^{n''}}{t^{m-d+2}}\label{3pjjo}
\eea
where in (\ref{3pjoo}) we defined ${\bf t}= {\bf z'_{13}}-{\bf z'_{23}}$,
while in (\ref{3pjjo}) we defined ${\bf t}= {\bf z'_{21}}-{\bf z'_{31}}$.

Although we have worked with a generic $m$, the $7d$ gauged sugra action
determines it exactly in units of $AdS$ radius.
In the sugra action, the 
purely gravitational piece is $-1/2\int (R+\frac{15}{2}m^2)$, whereas for 
unit $AdS_7$ the gravitational contribution is $\int (R+30)$. This implies
$m=2$. 

Finally, there is one more check on our correlators. The $<J\cal{OO}>$ 
correlator must obey the Ward identity
\bea
&&\frac{\partial}{\partial {\bf z_3}^p}\langle J_p^{CD}({\bf z_3}) 
{\cal O}_{ijk}^A ({\bf z_1}){\cal O}_{lmn}^B ({\bf z_2})\rangle
=g\delta^{CD}_{AE}\delta({\bf z_{13}})\langle
O_{ijk}^E ({\bf z_1})O_{lmn}^B ({\bf z_2})\rangle\nonumber\\
&&+g\delta^{CD}_{BE}\delta({\bf z_{23}})\langle
O_{ijk}^A ({\bf z_1})O_{lmn}^E ({\bf z_2})\rangle
\label{ward}
\eea
Hence, the coefficients of $<J\cal{OO}>$ and $<\cal{OO}>$ correlators are 
interdependent. To simplify our calculations, we will verify the Ward 
identity for ${\bf z_1}=0$, ${\bf z_2}\equiv{\bf y}$ at infinity and 
${\bf z_3}\equiv {\bf x}$ at finite distance.
Substituting the 3-point function (\ref{3pjoo}) into the l.h.s. of (\ref{ward})
we get (for $m=2$)
\bea
&&\frac{\partial}{\partial x^p}\langle J_p^{CD}({\bf x}) 
{\cal O}_{ijk}^A (0){\cal O}_{lmn}^B ({\bf y})\rangle=
4 g m {c_1}^2 c_2\frac{\pi^3}
{10(m+3)(m+2)(m+1)}\delta^{AB}_{CD}\nonumber\\
&&J_{ll'}({\bf y})J_{mj}({\bf y})J_{nk}({\bf y})y^{-2(m+3)}\frac{\partial}
{\partial x^p}\left(\frac{x_{l'}x_i x_p}{x^8}\right)  
\eea
Further using that
\be
\frac{\partial}{\partial x^p}\left(\frac{x_{l'}x_i x_p}{x^8}\right)  =
\frac{\pi^3}{6}\delta({\bf x})\delta_{l'i}
\ee 
and by comparing with the r.h.s. of (\ref{ward}) we confirm the identity.

\section{Conclusions}
$${}$$

In this paper we have studied $p$-forms in $AdS_{2p+1}$. 
We have constructed the propagators for $p$-forms with topological mass 
terms and topological kinetic terms in $AdS_{2p+1}$ (the latter being the 
case of self-dual forms in odd dimensions). We have also found that the 
basis for maximally symmetric bitensors previously found by Allen and 
Jacobson is not complete is some situations, and found the missing bitensors.

For the $AdS-CFT$ correspondence, we have 
analyzed the case of $AdS_7$ - $6d$ (2,0) $CFT$ (similar computations go 
through for $AdS_5-4d SYM$). We have computed the two point 
function of the ``self-dual'' 3-form by adding a boundary term to the 
$7d$ gauged sugra action. We have also computed 3-point functions of 
two 3-forms and a gauge field, and two gauge fields and a 3-form by using 
methods similar to the ones of Freedman et al. \cite{fmmr}. 
It is the hope of the authors that this calculation
can shed some light on the $6d$ (2,0) $CFT$.

 {\bf Acknowledgements:} The authors would like to acknowledge 
useful conversations with 
Joe Polchinski, Daniel Z. Freedman, Asad Naqvi, Peter van Nieuwenhuizen, 
Martin Ro\v cek, and Radu Roiban. The work of I.B. was 
supported in part by NSF grant PHY97-22022 and and the work of H.N. and D.V. 
by NSF grant 9722101.


\appendix1
\setcounter{subsection}{0}
\subsection{Conventions and useful identities involving the chordal distance}

We used the symbols $*$ and $d$ 
acting on forms, or on the unprimed indices of propagators, with the 
following normalization:  
\be
(*A)^{\mu_1 ... \mu_p} =  {1 \over (d-p)!} \epsilon^{\mu_1\mu_2 ... 
\mu_{d+1} } A_{\mu_{p+1}...\mu_{d+1}}  \eqno(3)$$
$$(dA)_{\mu_1 ... \mu_{p+1}} = {1\over p!} D_{[\mu_{1}}A_{\mu_{2}...
\mu_{p+1}]}
\ee
where the square brackets denote antisymmetrization of all unprimed indices.

In the computations the following identities were useful:
\bea
&&\p_{\mu} \p_{\nu'}u = {-1 \over z_0 w_0}[\delta_{\mu\nu'}+{(z-w)_\mu 
\delta_{\nu'0}\over w_0}+       {(w-z)_\nu' \delta_{\mu0}\over z_0} - u
\delta_{\mu 0}\delta_{\nu' 0}] \\
&&\p_{\mu} u = {1 \over z_0}[(z-w)_{\mu}/w_0-u \delta_{\mu 0}]  \\
&&\p_{\nu'} u = {1 \over w_0}[(w-z)_{\nu'}/z_0-u \delta_{\nu' 0}] \\
&& D^{\mu}\p_{\mu}u=(d+1)(u+1)          \\                       
&&      \p^{\mu}u\  \p_{\mu}u = u(u+2)          \\
 &&      D_{\mu}\p_{\nu}u       =g_{\mu\nu}(u+1)        \\
&&      (\p^{\mu}u)( D_{\mu}\p_{\nu}\p_{\nu'} u) = \p_{\nu}u \p_{\nu'} u \\
&&(\p^{\mu}u)( \p_{\mu}\p_{\nu'}) u =( u+1) \p_{\nu'} u                 \\
&&       D_{\mu}\p_{\nu}\p_{\nu'} u =           g_{\mu\nu}  \p_{\nu'} u
\eea


\subsection{Limits}
In this paragraph we present the arguments leading to the various limits
taken in the body of this paper.
First, we consider the limit
\be
\lim_{\epsilon\rightarrow 0}\frac{\epsilon^{2m}}{({\bf x}^2+\epsilon^2)^{3+m}}
=\frac{1}{m(m+1)(2+m)}Area(S_6)\delta^6(x)\label{unu}
\ee
where $Area(S_6)$ is the area of a 6d sphere of radius 1. 
To derive this limit, use a scaling argument to show that 
$\int d^6x \epsilon^{2m}/
({\bf x}^2+\epsilon^2)^{3+m}$ (with positive integrand) is independent 
of $\epsilon$. It is also obvious 
that as long as the denominator is nonzero, the limit is vanishing. Thus,
the limit is proportional to a delta function. The proportionality constant
is determined by evaluating the integral $\int d^6x \epsilon^{2m}/
({\bf x}^2+\epsilon^2)^{3+m}$ in spherical coordinates.

The limit 
\be
\lim_{\epsilon\rightarrow 0}\frac{\epsilon^{2m+2}}{({\bf x}^2+
\epsilon^2)^{4+m}}
=\frac{1}{(m+1)(m+2)(m+3)}Area(S_6)\delta^6({\bf x})\label{doi}
\ee
follows from (\ref{unu}), by redefining $m\rightarrow m+1$.

Finally, we derive the limit
\be
\lim_{\epsilon\rightarrow 0}\frac{\epsilon^{2m} {\bf x}^2}
{({\bf x}^2+\epsilon^2)^{4+m}}=\frac{3}{m(m+1)(m+2)(m+3)}Area(S_6)
\delta^6({\bf x})\label{trei}
\ee
For the proof of this limit we consider again the integral $\int d^6x
\epsilon^{2m} {\bf x^2}/({\bf x}^2+\epsilon^2)^{4+m}$. By the same
scaling argument, it is clear that the result is independent of $\epsilon$.
The limit is zero, as long as the denominator is non-zero. By adding and 
subtracting $\epsilon^2$ in the limit numerator and by making use  
of (\ref{unu},\ref{doi}), we determine the 
proportionality constant on the r.h.s. 
(\ref{trei}). 

As a consequence of (\ref{trei}) we obtain the following result:
\bea
&&\lim_{\epsilon \rightarrow 0} \int d^6x \epsilon^{2m} 
\frac{{\bf x}^i {\bf x}^j}{({\bf x}^2+\epsilon^2)^{4+m}} S_{jkl}({\bf x+w})
\nonumber\\&=&
\lim_{\epsilon\rightarrow 0} \int_{S_6} d\Omega \int_0^\infty dr r^5
\epsilon ^{2m} \frac{r^2}{(r^2+\epsilon^2)^{4+m}} {\bf n}_i {\bf n}_j 
S_{jkl}({\bf x+w})\nonumber\\
&=& 
\frac{6}{m(m+1)(m+2)(m+3)} \int_{S_6} d\Omega\int_0^\infty dr \delta(r) 
{\bf n}_i {\bf n}_j S_{jkl}({\bf x+w})\nonumber\\
&=&\frac{1}{2m(m+1)(m+2)(m+3)}Area(S_6)S_{ikl}({\bf w})
\label{patru}
\eea
where ${\bf n}_i={\bf x}_i/|{\bf x}|$.
\subsection{Integrals}

In this paragraph we record the integrals needed in the computation of 
3-point functions.
We begin (for completeness) with the integral (\cite{fmmr})
\bea
&&I(a,b,c)({\bf t})\equiv\int d^7w \frac{(w^0)^a}{(w-{\bf t})^{2b} w^{2c}}
\nonumber\\
&&=C(a,b,c)|{\bf t}|^{7+a-2b-2c}\label{inti}
\eea
where the constant $C(a,b,c,)$ is:
\bea
C(a,b,c)&=&\frac{\pi^3}{2}\frac{\Gamma({\frac{a}{2}+\frac 12}) \Gamma(
b+c-3-\frac{a}{2} -\frac{1}{2})}{\Gamma(b)\Gamma(c)}\nonumber\\
&&\frac{\Gamma(\frac 12 +\frac a2 +3-b)\Gamma(\frac 12 +\frac a2 +3-c)}
{\Gamma(7+a-b-c)}
\eea
Then, the following integrals can be obtained from (\ref{inti}) by 
differentiating with respect to the vector ${\bf t}$:
\bea
&&J_{i}(a,b+1,c)\equiv\int d^7w \frac{(w^0)^a w_i}{(w-{\bf t})^{2(b+1)} w^{2c}}
\nonumber\\
&&=\frac{\partial_i I(a,b,c)}{2b}+t_i I(a,b+1,c)\label{intj1}\\
&&J_{i_1i_2}(a,b+2,c)\equiv\int d^7w \frac{(w^0)^a w_{i_1}w_{i2}}
{(w-{\bf t})^{2(b+2)} w^{2c}}\nonumber\\
&&=\frac{\partial_{i_1}\partial_{i_2}I(a,b,c)}{2^2 b(b+1)}
+t_{i_1} t_{i_2} I(a,b+2,c)+\frac{\delta_{i_1 i_2}I(a,b+1,c)}{2(b+1)}
\nonumber\\
&&+\frac{(t_{i_1}\partial_{i_2}+1\;more)I(a,b+1,c)}{2(b+1)}\label{intj2}\\
&&J_{i_1i_2i_3}(a,b+3,c)\equiv\int d^7w \frac{(w^0)^a w_{i_1}w_{i2}w_{i_3}}
{(w-{\bf t})^{2(b+3)} w^{2c}}\nonumber\\
&&=\frac{\partial_{i_1}\partial_{i_2}\partial_{i_3}I(a,b,c)}{2^3 b(b+1)(b+2)}+
t_{i_1}t_{i_2}t_{i_3}I(a,b+3,c)
\nonumber\\&&+\frac{(t_{i_1}t_{i_2}\partial_{i_3}+2\;more)
I(a,b+2,c)}{2(b+2)}+\frac{(t_{i_1}\partial_{i_2}\partial_{i_3}+2\;more)
I(a,b+1,c)}{2^2(b+1)(b+2)}
\nonumber\\&&+\frac{(\delta_{i_1i_2}t_{i_3}+2\;more)
I(a,b+2,c)}{2(b+2)}+\frac{(\delta_{i_1i_2}\partial_{i_3}I(a,b+1,c)+2\;more)
}{2^2(b+1)(b+2)}\nonumber\\
\label{intj3}
\eea

\end{document}